\documentclass[10pt,twocolumn,letterpaper]{article}

\usepackage[utf8]{inputenc}
\usepackage[T1]{fontenc}
\usepackage[margin=0.75in,top=0.75in,bottom=1in]{geometry}
\usepackage{graphicx}
\usepackage{hyperref}
\usepackage{url}
\usepackage[english]{babel}
\usepackage{adjustbox}
\usepackage{tabularx}
\usepackage{booktabs}
\usepackage{makecell}
\usepackage{subcaption}
\usepackage{amsmath}
\usepackage{amssymb}
\usepackage{titlesec}
\usepackage{abstract}
\usepackage{listings}
\usepackage{xcolor}
\usepackage{parskip}
\PassOptionsToPackage{hyphens}{url}
\usepackage{hyperref}

\titleformat{\section}{\normalfont\bfseries\scshape}{\Roman{section}.}{0.5em}{}
\titleformat{\subsection}{\normalfont\itshape}{\Alph{subsection}.}{0.5em}{}
\titlespacing*{\section}{0pt}{8pt}{4pt}
\titlespacing*{\subsection}{0pt}{5pt}{2pt}

\makeatletter
\renewcommand{\@biblabel}[1]{[#1]}
\renewcommand{\thebibliography}[1]{%
  \section*{\textbf{References}}
  \small
  \list{\@biblabel{\@arabic\c@enumiv}}{%
    \settowidth\labelwidth{\@biblabel{#1}}%
    \leftmargin\labelwidth
    \advance\leftmargin\labelsep
    \usecounter{enumiv}%
    \let\p@enumiv\@empty
    \renewcommand\theenumiv{\@arabic\c@enumiv}}%
  \sloppy\clubpenalty4000\widowpenalty4000%
  \sfcode`\.\@m}
\makeatother

\title{\LARGE \textbf{A Complementary Visualisation Suite for Empirical Performance Analysis: Tempographs, Histograms, Ridgeline Plots, Stacked Bar Charts, and Combination Charts Applied to Beethoven's Piano and Cello Sonatas}}

\author{Dr Ignasi Sole \quad \texttt{ignasiphd@gmail.com} \quad \today}

\begin{document}

\maketitle
\thispagestyle{empty}
\pagestyle{empty}


\begin{abstract}
The choice of visualisation in empirical performance analysis is not a neutral presentation decision but an analytical one: different graphical forms reveal different features of the same dataset, and reliance on any single type systematically conceals what the others expose. This paper presents and argues for a suite of five complementary visualisation tools; tempographs, histograms with spline-smoothed probability density functions, ridgeline plots, stacked bar charts, and combination charts. These are applied to bar-level beats-per-minute data from recordings of Beethoven's five piano and cello sonatas (Op.~5 Nos.~1 and~2; Op.~69; Op.~102 Nos.~1 and~2) spanning 1930--2012. Each tool is described formally, its analytical properties characterised, its implementation detailed in working Python and MATLAB code, and its specific contribution demonstrated on a worked example using two recordings of Op.~5 No.~1 (Casals/Horszowski 1930--39 and Isserlis/Levin 2012) separated by eight decades. A five-panel composite figure applies all five tools to the same two recordings simultaneously, making the complementarity argument concrete: the tempograph reveals moment-to-moment structural parallels invisible in aggregate statistics; the spline-smoothed histogram exposes bimodality and secondary peaks suppressed by binning artefacts; the ridgeline plot positions both recordings within the full distributional space; the stacked bar chart shows divergent sectional pacing concealed by identical movement means; and the combination chart integrates mean tempo, variability, and historical reference marks in a single view. The spline-CDF smoothing method, applied to histogram data via cubic spline interpolation with zero-slope boundary conditions, is presented as a novel contribution to the performance analysis toolkit. Full implementation code is publicly available.
\end{abstract}


\section{Introduction}

Empirical performance analysis has developed a growing vocabulary of visualisation tools since Bowen's foundational work established scatter plots, bar graphs, and tempo maps as instruments for making interpretive decisions measurable~\cite{c1}. The CHARM-era studies of Cook and Leech-Wilkinson extended this vocabulary with tempographs, timescapes, and correlation networks~\cite{c2, c3}. Yet the field has been slower to adopt statistical distribution tools, such as histograms, kernel density estimates and ridgeline plots, that are standard in adjacent quantitative disciplines, and slower still to ask the meta-question that motivates this paper: what analytical consequences follow from choosing one visualisation type over another?

The answer is that the consequences are substantial. A bar chart of mean BPM per recording compresses each performance to a single number. It tells you nothing about whether that mean is the product of a consistently metronomic interpretation or a wildly variable one with a mathematically identical average. A tempograph reveals the second reading's variability in vivid detail but becomes unreadable when more than five recordings are overlaid. A ridgeline plot solves the overplotting problem for twenty or more recordings simultaneously but collapses the time dimension the tempograph preserves. A stacked bar chart recovers the time dimension at the level of formal sections rather than individual bars; a different resolution entirely. A combination chart integrates multiple summary statistics into one view but at the cost of compressing each individual dimension.

Each visualisation type is a projection of high-dimensional performance data onto a two-dimensional plane. Each projection preserves certain features and discards others. A researcher who uses only one type is, by the logic of projection, systematically blind to the features the other types would reveal. This paper's argument is that richest performance analysis uses the full suite, each tool applied where its specific strengths are relevant, and that understanding the analytical properties of each tool (what it shows, what it hides, and why), is itself a methodological contribution.

The paper is structured as follows. Section~II reviews existing visualisation practice in performance analysis and identifies gaps that the suite addresses. Section~III presents the corpus and data pipeline. Sections~IV through~VIII treat each of the five tools in turn: formal description, analytical properties, implementation, and worked example. Section~IX demonstrates the suite as a whole on a single pair of recordings. Section~X discusses limitations and the scope of applicability. Section~XI concludes.


\section{Existing Practice and Gaps}

\subsection{The Founding Vocabulary}

Bowen's 1996 study established the principal visualisation forms that remain dominant in empirical performance analysis~\cite{c1}. His scatter plots of performance duration against average tempo made flexibility visible as deviation from a regression line. His bar-by-bar tempo maps (the precursors of what are now called tempographs), captured rubato and structural tempo modulation at a resolution that aggregate statistics cannot achieve. His observation that overall duration is shaped less by average tempo than by internal tempo modulation~\cite{c1} is itself a visualisation-derived insight: it could not have been reached without a tool that distinguishes within-performance variation from between-performance variation.

Cook's work through CHARM extended these foundations in two directions. His tempograph analysis of Chopin Mazurkas in over 1,700 recordings~\cite{c2} demonstrated the scalability of bar-level timing analysis to large corpora. His use of timescapes and correlation networks showed that it is possible to visualise not just individual performances but relationships between them; a step toward the ecological mapping of repertoire traditions. Leech-Wilkinson's portamento studies~\cite{c3} applied scatter plots with regression lines to show long-term decline in expressive devices, demonstrating that visualisation can make eight decades of stylistic change legible in a single figure.

\subsection{What Is Missing}

What the existing literature lacks is a systematic treatment of the complementarity problem: how should different visualisation types be selected and combined to make different aspects of the same dataset visible? The implicit assumption in most studies is that one well-chosen visualisation type is sufficient. Bowen uses scatter plots and tempo maps. Cook uses tempographs and timescapes. Leech-Wilkinson uses scatter plots and duration tables. None of these studies applies multiple visualisation types to the same data and asks what each reveals that the others do not.

Three specific gaps are addressed by the suite proposed in this paper. First, the absence of statistical distribution tools from the musicological repertoire. Histograms, kernel density estimates, and ridgeline plots are standard in data science, psychology, and economics, but remain rare in performance analysis despite being directly applicable to bar-level tempo data. Second, the absence of a principled smoothing method for tempo histograms. Raw histograms are sensitive to bin width and boundary placement; a smoothed PDF derived from the empirical CDF addresses this without requiring parametric assumptions. Third, the absence of multi-variable composite visualisations that integrate tempo level, tempo variability, and historical benchmarks in a single figure, making it possible to identify interpretive profiles rather than individual statistics.


\section{Corpus and Data}

The corpus consists of recordings of the first movement of Beethoven's Op.~5 No.~1 cello sonata, used throughout as the worked example, alongside the full corpus of second and third movements across all five sonatas used for broader illustration. Recordings span 1930--2012. Performers include Pablo Casals, Emanuel Feuermann, Pierre Fournier, Jacqueline du~Pr\'e, Mstislav Rostropovich, Yo-Yo Ma, Steven Isserlis, Pieter Wispelwey, Anner Bylsma, and others totalling 22 recordings of Op.~5 No.~1.

Tempo data were collected using the manual bar-by-bar protocol described in Sole~\cite{c4}: a cumulative lap-timer procedure that yields millisecond-resolution timestamps for every barline, from which bar-level BPM values are computed as $\mathrm{BPM}_i = n_i \times 60 / \Delta t_i$, where $n_i$ is the number of beats in bar $i$ and $\Delta t_i$ is the bar's duration in seconds. The complete dataset is publicly available~\cite{c5}. The full implementation code for all five visualisation types is available in the same repository under \texttt{Coded Graphs/}.

The worked comparison throughout Sections~IV--IX uses two recordings of Op.~5 No.~1 first movement: Casals/Horszowski (1930--39) and Isserlis/Levin (2012), chosen for their 80-year separation, their membership in different $k$-means tempo clusters (slow and mid-range respectively, as established in Sole~\cite{c6}), and their contrasting pedagogical lineages.


\section{Tempographs}

\subsection{Description and Analytical Properties}

A tempograph is a line chart in which the horizontal axis carries bar index and the vertical axis carries BPM, producing a continuous tempo curve that traces the performer's moment-to-moment pacing decisions across the movement. It is the only tool in the suite that preserves the time dimension in full: every ritardando, every structural tempo contrast, every phrase-boundary relaxation is legible as a deviation from the surrounding tempo level, and its precise bar location is retained.

The tempograph's primary strength is close reading. By aligning the tempo curve with score landmarks (section boundaries, thematic returns, cadential moments), the analyst can correlate expressive timing choices directly with the musical text. Overlaying two or more tempographs in a single figure enables direct comparison of those choices at specific moments rather than in aggregate.

Its primary limitation is scalability. Beyond five overlaid curves, a single tempograph becomes visually crowded and individual curves lose distinctiveness. Two complementary formats address this: focused tempographs of up to five recordings over spans of up to 100 bars for close comparative analysis, and small-multiple grids of one tempograph per recording for full-corpus overview.

\subsection{Implementation}

Tempographs were generated in Google Sheets for rapid iteration and in Python/Matplotlib for publication-quality output. In Sheets, bar indices are placed in column A and per-recording BPM values in adjacent columns; Insert~$\rightarrow$~Chart produces a line chart configurable via the Chart Editor. In Python:

\begin{lstlisting}[language=Python]
import matplotlib.pyplot as plt
import numpy as np

bars = np.arange(1, len(bpm_casals) + 1)
fig, ax = plt.subplots(figsize=(12, 4))
ax.plot(bars, bpm_casals, color='#2166ac',
        lw=0.9, label='Casals (1930-39)')
ax.plot(bars, bpm_isserlis, color='#d6604d',
        lw=0.9, label='Isserlis (2012)')
ax.set_xlabel('Bar number')
ax.set_ylabel('Tempo (BPM)')
ax.legend()
\end{lstlisting}

Section shading is added via \texttt{ax.axvspan(start, end, alpha=0.15)} for each formal division, with landmark labels placed at the top of the shaded region.

\subsection{Worked Example}

Panel~(a) of Fig.~\ref{fig:five_panel} shows the tempograph for Casals (1930--39, blue) and Isserlis (2012, red) across all 400 bars of the Op.~5 No.~1 first movement. The figure's most striking feature is the structural parallelism of the two curves despite their 80-year separation: both performers articulate the gear-change from introduction to Allegro at bar 35 with virtually identical abruptness; both dip expressively at bars 72 and 159 (structural transitions within the Allegro); and both produce the same coda explosion at bars 368--383, reaching 217--238~BPM. Isserlis runs 5--8~BPM faster through most of the Allegro, but the \emph{shape} of the tempo trajectory (where each performer breathes, pushes, and relaxes), is structurally congruent. This congruence is the tempograph's unique contribution: it is invisible in any aggregate statistic and fully visible here.


\section{Histograms with Spline-Smoothed PDFs}

\subsection{Description and Analytical Properties}

A histogram partitions the BPM range into discrete intervals and displays the frequency of bars falling into each. Applied to the complete bar-level BPM sequence of a single recording, it reveals the statistical distribution of tempo within that performance: a narrow, peaked histogram indicates metronomic consistency; a broad histogram indicates flexibility; a bimodal histogram indicates a performance with two structurally distinct tempo zones.

The histogram's principal limitation is sensitivity to binning: different bin widths and boundary placements produce different visual impressions of the same underlying distribution. Subtle features (secondary peaks, asymmetric tails), may appear or disappear depending on binning choices. A smoothed probability density function (PDF) derived from the empirical cumulative distribution function (CDF) addresses this without imposing a parametric model.

\subsection{The Spline-CDF Smoothing Method}

The smoothing procedure is as follows. Let $\{b_k\}_{k=1}^{K}$ be the histogram bin counts and $\{c_k\}$ their midpoints. The empirical CDF is computed by:

\begin{equation}
F_k = \frac{\sum_{j=1}^{k} b_j}{\sum_{j=1}^{K} b_j}, \quad k = 1, \ldots, K.
\label{eq:ecdf}
\end{equation}

Boundary points $(c_0 - \varepsilon,\, 0)$ and $(c_K + \varepsilon,\, 1)$ are prepended and appended to enforce the correct asymptotic behaviour. A cubic spline $S(x)$ is then fitted to the extended sequence $\{(c_k, F_k)\}$ with zero-slope boundary conditions:

\begin{equation}
S'(c_0 - \varepsilon) = 0, \qquad S'(c_K + \varepsilon) = 0.
\label{eq:bc}
\end{equation}

The boundary conditions enforce that $S(x)$ flattens at the data's minimum and maximum, preventing the common kernel density estimation (KDE) artefact of probability mass appearing outside the observed range~\cite{c7}. The smoothed PDF is then recovered by differentiating the spline:

\begin{equation}
\hat{f}(x) = S'(x) = \frac{dS}{dx},
\label{eq:pdf}
\end{equation}

evaluated on a fine grid of 400 points spanning the data range. The result is a continuous, non-negative density curve that traces the histogram's shape without bin-boundary artefacts.

This method differs from standard KDE in that it smooths the \emph{cumulative} distribution rather than placing a kernel at each individual data point. For discrete bar-level tempo data, where adjacent bars often share identical BPM values due to the measurement resolution, this cumulative approach naturally pools neighbouring observations and avoids the artificial spikes that point-centred KDE produces at repeated values.

\subsection{Implementation}

Histograms and spline PDFs were generated in Python using SciPy's \texttt{CubicSpline}:

\begin{lstlisting}[language=Python]
import numpy as np
from scipy.interpolate import CubicSpline

def spline_pdf(data, n_bins=28):
    counts, edges = np.histogram(data,
                    bins=n_bins, density=False)
    centres = 0.5 * (edges[:-1] + edges[1:])
    cdf = np.cumsum(counts).astype(float)
    cdf /= cdf[-1]
    # Add boundary points with zero slope
    x = np.concatenate(
        [[edges[0]-1e-3], centres, [edges[-1]+1e-3]])
    y = np.concatenate([[0.], cdf, [1.]])
    cs = CubicSpline(x, y,
         bc_type=((1, 0.0), (1, 0.0)))
    x_fine = np.linspace(edges[0], edges[-1], 400)
    pdf = np.maximum(cs(x_fine, 1), 0)
    return counts, edges, centres, x_fine, pdf
\end{lstlisting}

A small random jitter (uniform $\pm 0.5$~BPM) is added to each measurement before binning to prevent artificial spikes at repeated values~\cite{c8}.

\subsection{Worked Example}

Panels~(b) of Fig.~\ref{fig:five_panel} show two histograms with superimposed spline PDFs for Casals (left) and Isserlis (right). Both distributions are strongly bimodal, with a low-BPM cluster around 30--45~BPM (the introduction) and a dominant high-BPM cluster at 130--160~BPM (the Allegro). The spline PDF reveals a feature the raw histogram partly obscures: a secondary shoulder in the Casals distribution around 85--100~BPM, corresponding to his expressive dipping at transitional moments within the Allegro. The Isserlis distribution is considerably narrower in its Allegro peak (centred at approximately 149~BPM, standard deviation 23.4~BPM for the Allegro section) compared to Casals (138.8~BPM, standard deviation 23.8~BPM). The nearly identical standard deviations indicate that both performers deploy the same \emph{degree} of internal flexibility, just at different overall tempo centres. This is a finding that the tempograph suggested and the histogram makes quantitatively precise.


\section{Ridgeline Plots}

\subsection{Description and Analytical Properties}

A ridgeline plot (informally called a Joy Plot~\cite{c9}) stacks multiple density curves vertically, each representing the tempo distribution of one recording. Each ridge is drawn using kernel density estimation~\cite{c7}, producing a smooth approximation of the underlying distribution. The curves are offset vertically to prevent overlap while sharing a common horizontal axis, so that the same BPM value aligns vertically through all ridges.

The ridgeline plot's specific advantage is the combination of individual-recording resolution with full-corpus simultaneity. A conventional histogram grid achieves the same individual resolution but requires more page space and makes direct visual comparison harder. An overlaid density plot achieves simultaneity but produces visual chaos beyond three or four recordings. The ridgeline format, by separating each distribution onto its own horizontal strip, avoids both problems~\cite{c10}.

Colour coding adds an analytical layer: a coolwarm palette mapping ridge colour to the recording's mean BPM allows historical tempo trends to be read directly from the visual structure of the figure. If the colour palette transitions systematically from cooler to warmer tones as one descends the stack (when recordings are arranged chronologically), that transition is direct visual evidence of a historical tempo trend.

The ridgeline plot's trade-off is the loss of the time dimension; it shows the distribution of BPM values but not where in the movement each value occurred.

\subsection{Implementation}

Ridgeline plots were generated in Python using Matplotlib and SciPy's \texttt{gaussian\_kde}, without relying on dedicated ridgeline libraries in order to retain full control over the rendering:

\begin{lstlisting}[language=Python]
from scipy.stats import gaussian_kde
import matplotlib.pyplot as plt
import numpy as np

recordings = [('Casals (1930)', bpm_c, '#2166ac'),
              ('Isserlis (2012)', bpm_i, '#d6604d')]
offsets = [0.55, 0.0]
scale = 4.5

fig, ax = plt.subplots(figsize=(10, 3))
for (label, data, col), offset in zip(
        recordings, offsets):
    kde = gaussian_kde(data, bw_method=0.07)
    x = np.linspace(0, 265, 800)
    y = kde(x) * scale + offset
    ax.fill_between(x, offset, y,
                    alpha=0.55, color=col)
    ax.plot(x, y, color=col, lw=2.0)
    ax.axvline(np.mean(data), color=col,
               lw=1.3, ls=':', alpha=0.85)
ax.set_yticks([])
ax.set_xlabel('Tempo (BPM)')
\end{lstlisting}

The bandwidth parameter \texttt{bw\_method=0.07} was selected by visual inspection to resolve the bimodal structure (introduction vs.\ Allegro) without over-smoothing. For large-corpus ridgeline plots with many recordings, Seaborn's \texttt{kdeplot} within a \texttt{FacetGrid} provides a more scalable implementation~\cite{c11}:

\begin{lstlisting}[language=Python]
import seaborn as sns
g = sns.FacetGrid(bpms_long,
    row='recording', hue='average',
    palette='coolwarm', aspect=6, height=0.8)
g.map(sns.kdeplot, 'bpm', fill=True,
      alpha=1, linewidth=1.5)
g.fig.subplots_adjust(hspace=-0.5)
g.despine(bottom=True, left=True)
\end{lstlisting}

\subsection{Worked Example}

Panel~(c) of Fig.~\ref{fig:five_panel} shows the two-recording ridgeline. Both ridges display the same bimodal structure as the histograms. The Casals ridge (upper, blue) has a wider Allegro peak and a more pronounced low-BPM introduction cluster. The Isserlis ridge (lower, red) has a taller, narrower Allegro peak shifted approximately 6--8~BPM to the right. The dotted vertical lines mark each recording's overall mean, making the tempo difference immediately legible without obscuring the distributional detail that the mean alone would suppress.

In a full-corpus application, this panel would contain 22 ridges rather than two, arranged chronologically and colour-coded by mean BPM. The horizontal alignment of ridge peaks across the stack would then reveal any historical drift in modal tempo, a trend that would be statistically estimable from the ridgeline but visually obvious before any regression is run.


\section{Stacked Bar Charts}

\subsection{Description and Analytical Properties}

The stacked bar chart is the only tool in the suite that reveals sectional pacing decisions: how each performer distributed total performance time across the formal divisions of the movement. Each bar represents one recording; each coloured segment within the bar represents one formal section's duration as a proportion of the total movement duration.

This tool recovers a dimension that tempographs, histograms, and ridgeline plots all collapse: the relative weight assigned to different structural regions. A performer who assigns disproportionate time to the development relative to the exposition is making a specific interpretive claim about where the movement's expressive weight lies; a claim visible in the stacked bar chart and invisible in every other tool in the suite. The stacked bar chart's trade-off is that it compresses all within-section tempo behaviour to a single duration value per section per recording.

\subsection{Implementation}

Section durations were computed by summing bar-level durations $\Delta t_i = 60 \times n_i / \mathrm{BPM}_i$ (in seconds) across all bars within each formal division. Section boundaries were defined by bar number from the score, shared across all recordings. Charts were generated in Google Sheets for rapid production and in Python/Matplotlib for the composite figure:

\begin{lstlisting}[language=Python]
import matplotlib.pyplot as plt
import numpy as np

sections = ['Introduction', 'Exposition',
            'Development', 'Recapitulation',
            'Coda']
colours = ['#d1e5f0','#fddbc7','#e0f3db',
           '#fee090','#f5f5f5']
bottoms = [0.0, 0.0]  # one per recording

for sec, col in zip(sections, colours):
    vals = [dur_casals[sec], dur_isserlis[sec]]
    ax.bar(positions, vals, bottom=bottoms,
           color=col, edgecolor='white',
           label=sec)
    bottoms = [b + v for b, v in
               zip(bottoms, vals)]
\end{lstlisting}

\subsection{Worked Example}

Panel~(d) of Fig.~\ref{fig:five_panel} shows section-level duration for both recordings. Casals's total movement duration is 905 seconds; Isserlis's is 851 seconds; a difference of 54 seconds, approximately 6\%. The exposition durations are closely matched (248~s vs.\ 219~s proportionally), and the introduction (100~s vs.\ 100~s) is essentially identical despite the slightly higher mean introduction BPM for Isserlis. The most striking difference lies in how each performer treats the recapitulation: Casals performs it at 224~s compared to Isserlis's 214~s, a difference consistent with the 5--8~BPM tempo divergence visible in the tempograph. The coda is proportionally identical (approximately 16 seconds each), confirming that both performers take the same approach to the movement's culminating gesture.

These sectional differences would not be visible in the tempograph alone (which shows the within-section curves but not the proportional allocation) nor in the histogram (which discards section identity entirely). The stacked bar chart makes them the primary visual feature.


\section{Combination Charts}

\subsection{Description and Analytical Properties}

The combination chart is the suite's most information-dense tool. It overlays three distinct data series in a single figure: mean BPM per recording (or per recording group) displayed as filled bars against a primary y-axis; tempo variability (standard deviation of bar-level BPM) displayed as a line plot against a secondary y-axis; and horizontal reference lines marking historical metronomic recommendations from Czerny, Moscheles, and Kolisch.

This multi-layer design allows the identification of \emph{interpretive profiles} rather than individual statistics. A recording that is fast, consistent, and close to Kolisch's recommendation is a different interpretive proposition from one that is slow, variable, and far below Czerny's; and the combination chart makes this difference visible as a compound visual signature rather than requiring the reader to cross-reference three separate figures.

The trade-off is precision on each individual measure: the dual-axis design compresses the range of both y-axes and can make small differences in variability less legible than a dedicated variability chart would. The combination chart is therefore a synthesis tool, appropriate for overview and hypothesis generation, rather than for fine-grained statistical analysis.

\subsection{Implementation}

Combination charts were generated in Google Sheets for corpus-level overviews and in Python for the composite figure. The dual-axis design uses Matplotlib's \texttt{twinx()} method:

\begin{lstlisting}[language=Python]
fig, ax1 = plt.subplots(figsize=(5, 4))
ax2 = ax1.twinx()

# Bars: mean BPM (Allegro only)
ax1.bar(positions, means, color=bar_colours,
        edgecolor=edge_colours, linewidth=1.2)

# Line: standard deviation
ax2.plot(positions, stds, 'o--',
         color='#555555', lw=1.5, ms=7)

# Historical reference lines
for bpm, label, colour, ls in [
    (160, 'Czerny', '#e41a1c', '--'),
    (160, 'Moscheles', '#ff7f00', ':'),
    (126, 'Kolisch', '#4daf4a', '-.')]:
    ax1.axhline(bpm, color=colour,
                lw=1.3, ls=ls, label=label)
\end{lstlisting}

\subsection{Worked Example}

Panel~(e) of Fig.~\ref{fig:five_panel} shows the combination chart for Casals and Isserlis. The Allegro mean for Casals is 138.8~BPM; for Isserlis, 144.4~BPM. Both lie well below the Czerny and Moscheles recommendation of 160~BPM, and above Kolisch's more moderate suggestion of 126~BPM. Standard deviations (shown as dots on the right axis) are nearly identical: 23.8~BPM for Casals and 23.4~BPM for Isserlis. This finding is significant and would not be evident from the tempograph or the histograms alone: despite the 5.6~BPM difference in average tempo, the two performers deploy the same \emph{degree} of expressive flexibility. The difference between them is a tempo-level choice, not a flexibility-style choice.

The historical reference lines contextualise both recordings within the broader debate about Beethoven's metronome markings. Both Casals and Isserlis fall between Kolisch and Czerny/Moscheles, consistent with the finding across the full corpus that performers systematically fall below Czerny's and Moscheles's recommendations while remaining above Kolisch's more conservative markings~\cite{c12}.


\section{The Suite Applied: Five Panels, Two Recordings}

\begin{figure*}[htbp]
  \centering
  \includegraphics[width=\textwidth,height=0.82\textheight,keepaspectratio]{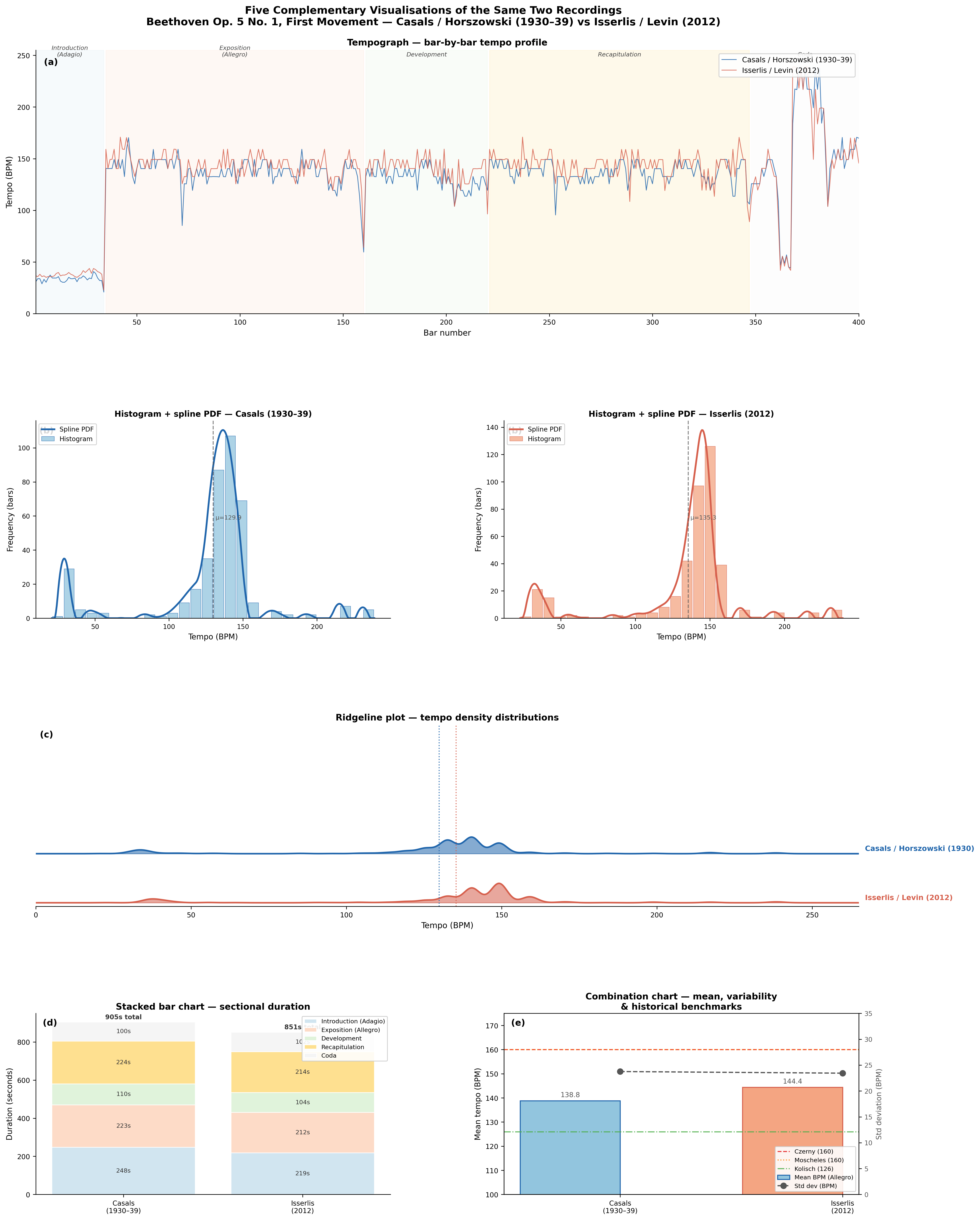}
  \caption{Five complementary visualisations applied to the same two recordings of Beethoven's \textit{Cello Sonata No.~1}, Op.~5, first movement: Casals/Horszowski (1930--39) and Isserlis/Levin (2012). (a)~Tempograph: bar-by-bar BPM across all 400 bars, with formal sections shaded. (b)~Histograms with spline-smoothed PDFs: tempo distributions for each recording, with spline-derived PDF in bold colour and mean indicated by dashed vertical line. (c)~Ridgeline plot: KDE-based density distributions for both recordings on a shared BPM axis. (d)~Stacked bar chart: sectional duration in seconds for each recording. (e)~Combination chart: Allegro mean BPM (bars, left axis), standard deviation (dots, right axis), and historical reference lines for Czerny, Moscheles, and Kolisch.}
  \label{fig:five_panel}
\end{figure*}

Fig.~\ref{fig:five_panel} assembles all five panels for the two recordings, making the complementarity argument concrete rather than abstract. Reading the figure from panel to panel is a demonstration of what each tool contributes and what it cannot show alone.

The tempograph~(a) reveals the most detailed picture: structural congruence between the two interpretations despite their 80-year separation, with Isserlis running consistently faster through the Allegro. It shows \emph{where} each performer is faster or slower and \emph{by how much} at every bar. What it cannot show is whether this 5--8~BPM difference is consistent across the full range of tempi the movement contains, or whether it is concentrated at certain tempo levels.

The histograms~(b) answer that question. Both distributions are bimodal, with the secondary low-BPM cluster corresponding to the introduction. The Isserlis peak is narrower and shifted rightward, confirming a consistently higher Allegro tempo and marginally greater metronomic consistency. The spline PDF for Casals reveals a secondary shoulder around 85--100~BPM that the raw histogram partly obscures, the expressive dipping at structural transition points that the tempograph shows moment-by-moment but whose aggregate frequency is only visible here.

The ridgeline~(c) positions both recordings in the same distributional space, making the tempo difference and the distributional shape differences simultaneously visible. In a full-corpus application this panel would contain 22 stacked ridges, immediately conveying the range of tempo traditions across the entire recording history of the sonata.

The stacked bar chart~(d) reveals a dimension invisible in all three preceding panels: Casals's total duration is 54 seconds longer than Isserlis's, and this difference is distributed across sections in a non-uniform way. The introduction durations are essentially identical despite the slightly different introduction BPM values. The exposition and recapitulation are proportionally similar. The aggregate duration difference is driven by the Allegro sections collectively, not by any single structural moment receiving special expressive expansion.

The combination chart~(e) integrates the key comparative statistics into a single view: both performers are substantially below Czerny's recommendation, Isserlis is 5.6~BPM faster, and their tempo variability is nearly identical. This last finding (that two performers separated by eight decades use the same standard deviation of tempo), is the combination chart's unique contribution to this worked example. It reframes the difference between the two recordings: not as a difference in expressive style (both are equally flexible) but as a difference in tempo centre (Isserlis has normalised upward by 5.6~BPM relative to Casals).

Taken together, the five panels construct a richer interpretive account of the difference between these two recordings than any single tool could produce. The methodological implication is direct: any study that applies only one visualisation type to this kind of data is systematically leaving information on the table.


\section{Discussion}

\subsection{Choosing the Right Tool for the Question}

The five tools in the suite are not interchangeable, and the choice among them should be driven by the research question at hand. Table~\ref{table:tool_selection} summarises the mapping between question types and appropriate tools.

\begin{table}[htbp]
\caption{Tool selection by research question}
\centering
\small
\begin{tabularx}{\columnwidth}{|X|X|}
\hline
\textbf{Research question} & \textbf{Appropriate tool(s)} \\
\hline
Where does performer A deviate from performer B at specific moments? & Tempograph \\
\hline
How consistently does a performer maintain their mean tempo? & Histogram + spline PDF \\
\hline
How does one recording's tempo distribution compare to twenty others? & Ridgeline plot \\
\hline
How did a performer distribute time across formal sections? & Stacked bar chart \\
\hline
How does mean tempo, variability, and historical compliance compare across recordings? & Combination chart \\
\hline
What has changed historically across decades of recordings? & All five, applied sequentially \\
\hline
\end{tabularx}
\label{table:tool_selection}
\end{table}

\subsection{The Spline-CDF Method: Scope and Limitations}

The spline-CDF smoothing method offers advantages over standard KDE for discrete, bounded performance data, but it is not universally superior. For datasets with very few observations (fewer than approximately 50 bars), the empirical CDF is too coarsely estimated for the spline to produce a reliable smooth. For datasets with continuous measurement (e.g., onset-detection output from automated tools), standard KDE may be preferable because the discrete-value clustering that motivates the CDF approach does not arise. The method is specifically suited to bar-level tempo data, where the measurement resolution is coarser than the underlying continuous tempo and where boundary effects (tempo values outside the observed range having zero probability) are analytically meaningful.

\subsection{Scalability}

The suite scales differently across tools. Tempographs scale well to 100 bars and five recordings but poorly to full movements and 22 recordings simultaneously (requiring the small-multiple grid format). Histograms and combination charts scale to any corpus size but provide only recording-level summaries. The ridgeline plot scales most efficiently to large corpora; 22 recordings in a single figure is readable, 50 begins to challenge legibility. The stacked bar chart scales to approximately 20 recordings before horizontal crowding requires selecting a representative subset.

For corpora larger than those considered here, automated tempo extraction may become necessary despite its limitations on historical polyphonic recordings. The visualisation suite is tool-agnostic with respect to the data source: the same five chart types apply whether the BPM values come from manual annotation or automated extraction, provided the data's reliability characteristics are documented.

\subsection{Reproducibility and Open Data}

All figures in this paper were generated with Python 3.10 using Matplotlib~\cite{c13}, Seaborn~\cite{c14}, SciPy~\cite{c15}, NumPy, and Pandas~\cite{c16}. The complete figure-generation script, including all data arrays, is available at the repository~\cite{c5} under \texttt{Coded Graphs/five\_panel\_figure.py}. This script is self-contained and produces the five-panel composite figure from the raw BPM arrays without requiring any external data files, enabling direct verification and adaptation by other researchers.


\section{Conclusion}

This paper has argued that the choice of visualisation in empirical performance analysis is an analytical decision with direct consequences for what can be known about a dataset, and has presented a suite of five complementary tools (tempograph, spline-smoothed histogram, ridgeline plot, stacked bar chart, and combination chart), that together make accessible the full range of information contained in bar-level tempo data.

The suite's central contribution is the complementarity argument, demonstrated concretely in the five-panel composite figure applying all tools to the same two recordings of Beethoven's Op.~5 No.~1. Each panel contributes findings that the others cannot: the tempograph reveals structural congruence across eight decades; the histograms quantify distributional shape and expose secondary peaks; the ridgeline plots the two recordings in distributional space; the stacked bar chart shows sectional pacing allocation; and the combination chart establishes that two performers with a 5.6~BPM tempo difference deploy identical degrees of expressive flexibility. No single tool yields all five findings; all five tools are needed to yield any one of them confidently.

The spline-CDF smoothing method for histogram PDFs is offered as a specific technical contribution: a principled, parameter-light approach to obtaining a smooth density estimate from discrete bar-level tempo data that avoids the bin-boundary sensitivity of raw histograms and the boundary-leakage artefacts of standard KDE.

The suite is transferable. Any performance corpus with bar-level or onset-level timing data, such as Chopin Mazurkas, Brahms symphonies, Bach unaccompanied suites and others, is a candidate for this analysis. The research question drives the tool selection; the tools drive the findings; and the findings, taken together across all five visualisation types, constitute a more complete account of interpretive practice than any single projection of the data could provide.


\end{document}